\documentclass[runningheads]{llncs}
\usepackage[T1]{fontenc}
\usepackage{graphicx}
\usepackage{booktabs}
\usepackage{placeins}
\usepackage{amsmath}
\usepackage{url}
\newcommand{\ParticipantCount}{25}
\newcommand{\ControlledPairingsPerGenuineVideo}{24}
\newcommand{\ControlledStudyUsageRecordingCount}{150}

\newcommand{\ControlledComposedEvaluationVideoCount}{3600}

\newcommand{\EnrollmentRecordingCount}{100}

\newcommand{\InTheWildParticipantCount}{24}
\newcommand{\InTheWildUsageRecordingCount}{121}
\newcommand{\InTheWildComposedEvaluationVideoCount}{2178}

\newcommand{\InTheWildPairingsPerGenuineVideo}{18}
\newcommand{\MinInTheWildRecordings}{4}
\newcommand{\MaxInTheWildRecordings}{8}
\newcommand{\MeanInTheWildRecordings}{5}
\newcommand{\ParticipantAgeRange}{20--64}
\newcommand{\ParticipantMeanAge}{39}
\newcommand{\MaleParticipantCount}{11}
\newcommand{\FemaleParticipantCount}{14}
\newcommand{\InTheWildAndroidParticipantCount}{6}
\newcommand{\InTheWildIPhoneParticipantCount}{18}

\begin{document}
    \title{Continuous Face Authentication on Mobile and Desktop Platforms: A Comparative Study}
    \titlerunning{Continuous Face Authentication on Mobile and Desktop Platforms}


    \author{
        Miriam Palmetshofer
        \and
        Rainhard Dieter Findling
        \and
        Marc Kurz
    }
    \authorrunning{M.\ Palmetshofer \and R.\ D.\ Findling \and M.\ Kurz}
    \institute{SAIL Department, University of Applied Sciences Upper Austria, Hagenberg}
    \maketitle              % typeset the header of the contribution
    \begin{abstract}
        Personal devices hold sensitive data and provide access to sensitive services. Conventional personal device authentication verifies users' identity only at the moment access is granted. An unlocked device may be accessed by an unauthorized person if the user stops using the device without locking it, or if another person takes over. Continuous authentication addresses this gap.
        This paper investigates how device type and usage conditions influence continuous mobile face authentication with an InsightFace-based approach with temporal trust decay. We evaluate the approach with mobile and desktop recordings with different head directions and lighting conditions. We also evaluate recordings from everyday mobile device use without predefined tasks.
        The results show that device type alone has little impact, while different usage conditions do have impact on the authentication performance. Results also show that everyday mobile device use is in general more challenging for continuous face authentication, where reduced face visibility, including occlusions and faces outside the camera viewport, is a main contributor to false rejections.

    \end{abstract}

    \section{Introduction}
    \label{sec:intro}
    The security of personal devices relies on authentication mechanisms that verify user identity. Traditional authentication methods, such as passwords, PINs, or fingerprint scans, verify identity only once at the beginning of a session, securing initial access but leaving devices vulnerable during active use~\cite{patelContinuousUserAuthentication2016}. Once authenticated, an unauthorized user who gains physical access to an unlocked device can freely access information without triggering any additional security checks. This is especially relevant for mobile devices, which often provide access to sensitive information such as banking services and private messages, and which are frequently used in public spaces and are more susceptible to theft or opportunistic access when left unattended while unlocked~\cite{mahbubActiveUserAuthentication2016}.

    Continuous authentication addresses this gap by repeatedly verifying user identity throughout a session rather than only at login~\cite{shendeDeepLearningBased2024}. Among various biometric modalities, face-based continuous authentication offers a passive and non-intrusive approach: it leverages the front-facing camera to periodically capture and verify the user's face during device usage. Unlike behavioral biometrics that require user interaction patterns, face-based authentication can operate transparently without disrupting the user experience~\cite{patelContinuousUserAuthentication2016}.

    Despite growing interest in continuous face authentication, several challenges
    remain~\cite{shendeDeepLearningBased2024}. Real-world deployment must handle changing lighting, non-frontal camera angles, user movement, and partial occlusions~\cite{smith-creaseyContextAwarenessImproved2018}. Although non-frontal camera angles and difficult lighting are recognized challenges~\cite{patelContinuousUserAuthentication2016,smith-creaseyContextAwarenessImproved2018}, prior work has not explicitly compared how an approach behaves in controlled scenarios that emphasize these conditions. Likewise, continuous face authentication has been studied on mobile and desktop platforms separately, but direct comparisons across these platforms are still lacking.

    This paper addresses this gap through a comparative evaluation of continuous face authentication on mobile and desktop platforms under three predefined recording scenarios. To conduct this evaluation, we implement\footnote{The implementation is publicly available for research and teaching purposes at \url{https://github.com/miriampalmetshofer/conFaceAuth}.} an approach using the InsightFace framework~\cite{guo2021samplecomputationredistributionefficient,dengArcFaceAdditiveAngular2022} for face detection and verification, combined with a temporal authentication decay logic adapted from~\cite{Altenhofer_17_ContinuousMobileFace}, and evaluate its performance on both mobile and desktop platforms through a controlled study and an in-the-wild study. Our evaluation addresses two key research questions:
    \begin{enumerate}
        \item How does continuous face authentication performance compare between mobile and desktop platforms?

        \item How do challenging lighting conditions and repeated head turns affect continuous face authentication performance across mobile and desktop platforms?
    \end{enumerate}

    \section{Related Work}

    Continuous authentication verifies user identity throughout a device usage session rather than only at login. Many biometric modalities have been explored for this purpose. Early work studied behavioral biometrics such as keystroke dynamics, mouse movement, and touch gestures~\cite{patelContinuousUserAuthentication2016}. More recent approaches use smartphone motion sensors such as accelerometers and gyroscopes to model gait and device handling~\cite{abuhamadAUToSenDeepLearningBasedImplicit2020,liSNNAuthSensorBasedContinuous2024}. While these methods can perform well, they depend on sufficient user activity. Multimodal approaches combine several biometric traits to improve robustness~\cite{damerMultibiometricContinuousAuthentication2016,fenuMultibiometricSystemContinuous2018,hintzeCORMORANTUbiquitousRiskAware2019}, but also increase complexity. Face-based continuous authentication offers a practical alternative because it can rely only on the front-facing camera.

    In face-based continuous authentication approaches, early work relied on handcrafted features such as Local Binary Patterns (LBP)~\cite{samangoueiFacialAttributesActive2017,smith-creaseyContextAwarenessImproved2018}, Histograms of Oriented Gradients (HOG)~\cite{segundoContinuous3DFace2013,pereraFaceBasedMultipleUser2019}, and Eigenfaces~\cite{niinumaSoftBiometricTraits2010}. These methods are less robust to changes in pose, lighting, and facial expression. More recent approaches use deep learning models pretrained on large-scale face recognition datasets. FaceNet~\cite{schroffFaceNetUnifiedEmbedding2015}, for example, learns compact embeddings that place similar faces close together and different identities farther apart. Several continuous authentication approaches build on such models by extracting embeddings and comparing them with enrollment data using similarity metrics or user-specific classifiers~\cite{keykhaieLightweightSecureFaceBased2023,ganidisastraIncrementalTrainingDeep2021,ozaActiveAuthenticationUsing2019}. Match-on-card approaches have also been explored for stronger data protection~\cite{findlingMobileMatchonCardAuthentication2018}.

    Face detection is another key component in continuous face authentication approaches. Early real-time approaches often used Viola--Jones~\cite{violaRapidObjectDetection2001}, which was efficient but struggled with non-frontal poses and changing illumination in mobile scenarios~\cite{findlingTowardsPanShotFaceUnlock2013}. Other work explored depth-based face detection~\cite{findlingRangeFaceSegmentation2013} and multi-perspective capture strategies such as pan-shot face unlock~\cite{findlingTowardsPanShotFaceUnlock2013}. Modern detectors such as MTCNN~\cite{zhangJointFaceDetection2016} and BlazeFace~\cite{bazarevskyBlazeFaceSubmillisecondNeural2019} are more robust while remaining lightweight enough for mobile deployment~\cite{keykhaieLightweightSecureFaceBased2023}. For verification, recent approaches typically use Euclidean distance or cosine similarity, while some use user-specific classifiers such as SVMs~\cite{keykhaieLightweightSecureFaceBased2023,hussainal-najiCABIoTContinuousAuthentication2022,ozaActiveAuthenticationUsing2019}.

    Continuous face authentication also requires a temporal decision logic, as single-frame decisions are sensitive to motion blur, occlusion, and brief head turns~\cite{crouseContinuousAuthenticationMobile2015,hussainal-najiCABIoTContinuousAuthentication2022}. Sliding-window methods aggregate recent authentication results to reduce sensitivity to occasional low-quality frames~\cite{crouseContinuousAuthenticationMobile2015,hussainal-najiCABIoTContinuousAuthentication2022}. Other approaches use decay-based functions to model authentication confidence changes over time~\cite{Altenhofer_17_ContinuousMobileFace}.

    Prior work also differs in how continuous face authentication approaches are evaluated. MOBIO~\cite{khouryBimodalBiometricAuthentication2014} and UMDAA-02~\cite{mahbubActiveUserAuthentication2016} are common datasets for mobile continuous authentication research, although both have availability restrictions. Many studies evaluate approaches mainly in controlled laboratory settings~\cite{pereraFaceBasedMultipleUser2019,ozaActiveAuthenticationUsing2019}, while naturalistic data collection remains less common~\cite{crouseContinuousAuthenticationMobile2015}. Imposter simulations vary from concatenated video segments~\cite{segundoContinuous3DFace2013} to live handover trials~\cite{crouseContinuousAuthenticationMobile2015}.

    Overall, prior work shows that face-based continuous authentication is feasible, but evaluation outside controlled settings remains limited. While difficult lighting and non-frontal camera angles are acknowledged as challenges~\cite{patelContinuousUserAuthentication2016,smith-creaseyContextAwarenessImproved2018}, prior work has not explicitly compared approach behavior in controlled scenarios that emphasize these conditions. Likewise, although continuous face authentication has been studied on both desktop~\cite{ganidisastraIncrementalTrainingDeep2021,jsDeepLearningBased2021} and mobile~\cite{keykhaieLightweightSecureFaceBased2023,samangoueiFacialAttributesActive2017} platforms, direct comparisons across these platforms are still lacking.

    \section{Approach}
    \label{sec:approach}

\begin{figure*}[tbh]
        \centering
        \includegraphics[width=\linewidth]{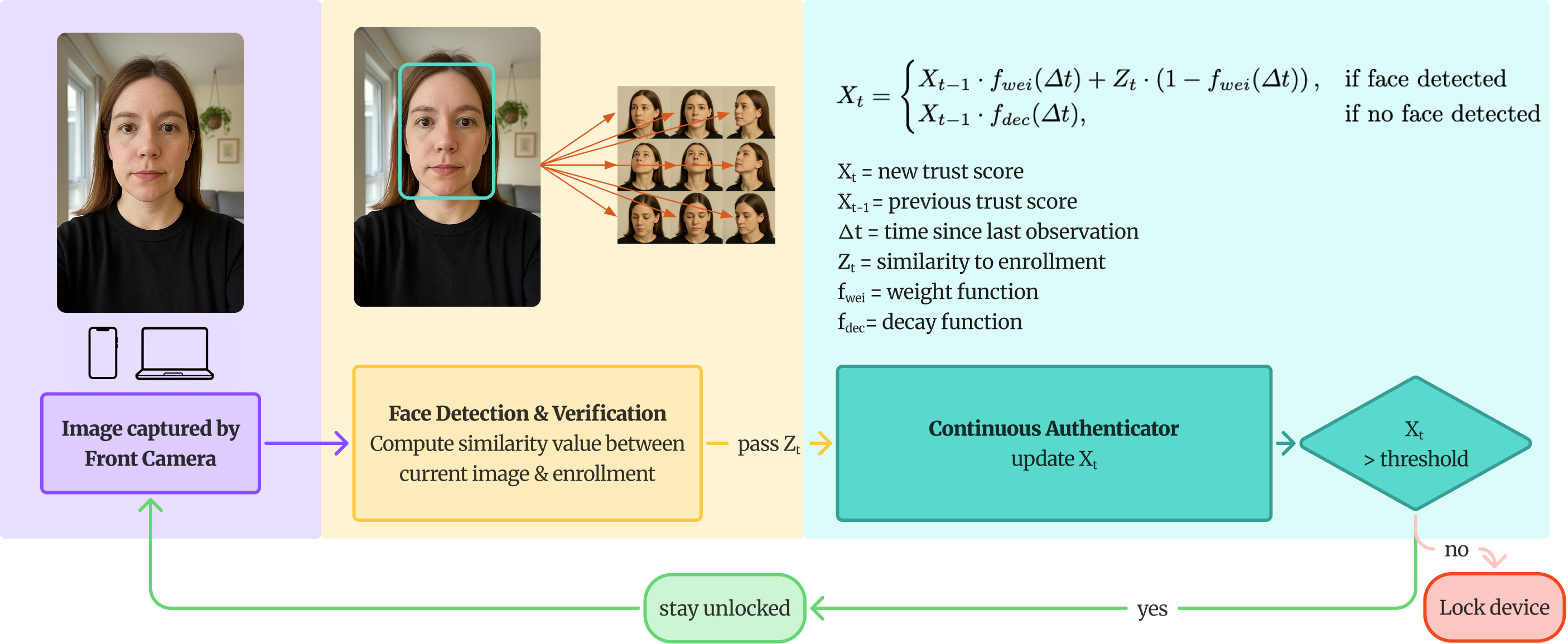}
        \caption{Overview of the continuous face authentication pipeline, from image capture and InsightFace-based verification to temporal trust updating and the authentication decision.}
        \label{fig:pipeline}
    \end{figure*}

    For our evaluation, we use a continuous face authentication approach that monitors user identity throughout a usage session by periodically analyzing frames from the device's front-facing camera (Fig.~\ref{fig:pipeline}). The approach maintains an internal trust score that reflects the likelihood the current user is the legitimate user. When this trust score drops below a predefined threshold, the device locks to prevent unauthorized access.1
    
    \subsection{Face Detection and Verification}
    \label{sec:face-detection-verification}

    We use InsightFace~\cite{guo2021samplecomputationredistributionefficient,dengArcFaceAdditiveAngular2022} for both face detection and embedding extraction. InsightFace performs detection and feature extraction in a single pass. Specifically, we use the \texttt{buffalo\_sc} model variant, which is the most compact option in the buffalo model family at 16 MB. This variant uses SCRFD\nobreakdash-500MF~\cite{guo2021samplecomputationredistributionefficient} for face detection and MobileFaceNet~\cite{chenMobileFaceNetsEfficientCNNs2018} for recognition. The recognition model generates 512\nobreakdash-dimensional L2\nobreakdash-normalized embedding vectors, and when multiple faces are detected in a frame, we select the largest one based on bounding box area, assuming it represents the primary device user.

    To determine whether an observed face belongs to the enrolled user, we compare the embedding extracted from the current frame with a set of $N$ enrollment embeddings obtained during enrollment. The enrollment consists of reference images captured under multiple head orientations (front, left, right, up, down), as described in Section~\ref{sec:enrollment}. For each incoming frame at time $t$, we first compute cosine similarities between the probe embedding $\mathbf{e}_{probe}$ and all $N$ enrollment embeddings $\mathbf{e}_{enroll}^{(i)}$ (Eq.~\ref{eq:cosine-similarities}).

    \begin{equation}
        \label{eq:cosine-similarities}
        s_i = \mathbf{e}_{probe} \cdot \mathbf{e}_{enroll}^{(i)}, \quad i = 1,\dots,N
    \end{equation}
    Since all embeddings are L2-normalized, the dot product corresponds to cosine similarity. Each similarity value $s_i$ ranges from $-1$ to $1$, where higher values indicate a closer match between the observed face and the enrolled user. To obtain a single similarity score for the frame, we compute the $p$-th percentile of these values using NumPy's default percentile function. This aggregation reduces the influence of poorly matching enrollment samples, for example due to unfavorable lighting or pose, while still considering multiple reference images (Eq.~\ref{eq:frame-similarity-score}).

    \begin{equation}
        \label{eq:frame-similarity-score}
        Z_t = \operatorname{percentile}_p(s_1, s_2, \dots, s_N) .
    \end{equation}
    The resulting $Z_t \in [-1,1]$ is the frame-level similarity score at time $t$, which the continuous authentication module uses to update its internal trust score.

    \subsection{Continuous Authentication}

    Our continuous authentication logic adapts the exponential time-weighting temporal decay approach of~\cite{Altenhofer_17_ContinuousMobileFace}. The preceding face detection and verification step produces one of two outcomes for each processed frame. If a face is detected, the similarity value $Z_t$ provides new evidence for updating the trust score. If no face is detected, no similarity value is available. The continuous authenticator handles these outcomes with the two branches. (Eq.~\ref{eq:trust-update}).

    \begin{equation}
        \label{eq:trust-update}
        X_t = \begin{cases}
                  X_{t-1} \cdot f_{wei}(\Delta t) + Z_t \cdot \left(1 - f_{wei}(\Delta t)\right), & \text{if a face is detected} \\
                  X_{t-1} \cdot f_{dec}(\Delta t), & \text{if no face is detected}
        \end{cases}
    \end{equation}
    $X_t$ is the trust score after processing frame $t$, and $X_{t-1}$ is the previous trust score. The score ranges from $-1$ to $1$, where higher values indicate greater confidence that the current user is legitimate.

    When a face is detected, the first branch uses the weight function $f_{wei}(\Delta t) = e^{-\Delta t / k_{weight}}$ to combine the previous trust score with the similarity value $Z_t$. The time constant $k_{weight}$ determines how quickly the influence of the previous score decreases. A larger value retains more of the previous score and makes individual observations less influential.
    For frames without a detected face, the second branch uses the decay function $f_{dec}(\Delta t) = e^{-\Delta t / k_{decay}}$ to reduce the previous trust score without a new similarity value. The time constant $k_{decay}$ determines the rate of this decrease. A larger value causes slower decay and allows more time before a temporary missed detection leads to a lock.

    Both functions use $\Delta t$, which measures the elapsed time since the previous trust-score update. This makes each update depend on elapsed time rather than on the number of processed frames. The functions therefore retain the same temporal interpretation when processing intervals differ or are irregular. Finally, the approach compares the current trust score $X_t$ to a predefined threshold $\tau$ to make authentication decisions (Eq.~\ref{eq:device-state-threshold}).

    \begin{equation}
        \label{eq:device-state-threshold}
        \text{Device State} =
        \begin{cases}
            \text{Unlocked} & \text{if } X_t \geq \tau \\
            \text{Locked} & \text{if } X_t < \tau
        \end{cases}
    \end{equation}
    The threshold $\tau$ and initial trust $X_0$ can be configured based on the desired balance between tolerance for brief genuine-user disruptions and rapid lockout of unauthorized users. When the trust score drops below the threshold, the device locks. Regaining access requires successful authentication through a non-continuous authentication mechanism, such as a PIN, password, or fingerprint.

    \subsection{Approach Configuration}

    Table~\ref{tab:pipeline-config} summarizes the configuration parameters used throughout the evaluation. We assume that continuous face authentication begins after a successful initial authentication step, such as fingerprint scanning or PIN entry. Accordingly, the initial trust score is set to $X_0 = 1.0$. The pipeline processes frames at 1\,Hz, so consecutive trust-score updates use $\Delta t = 1$\,s.

    The time constants $k_{weight} = 45$\,s and $k_{decay} = 60$\,s were empirically determined to balance responsiveness with tolerance for temporary quality degradation. To illustrate their practical impact, consider an initial trust of $X_0 = 1.0$, a threshold of $\tau = 0.5$, and updates at 1\,Hz. Under repeated weak face matches with fixed $Z_t = 0.12$, $k_{weight}$ being $20$\,s, $45$\,s, or $80$\,s leads to lockout times of 17\,s, 38\,s, and 68\,s, respectively. When no face is detected, $k_{decay}$ being $30$\,s, $60$\,s, or $100$\,s leads to lockout times of 21\,s, 42\,s, and 70\,s, respectively. Thus, increasing those constants makes the approach less reactive to weak matches or missing faces, while smaller constants lead to earlier lockout.

    \begin{table}[tbh]
        \centering
        \caption{Approach parameters used throughout the evaluation.}
        \label{tab:pipeline-config}
        \begin{tabular}{@{\extracolsep{\fill}}lc}
            \toprule
            \textbf{Parameter}                   & \textbf{Value} \\
            \midrule
            Authentication rate                  & 1\,Hz          \\
            Weight decay constant ($k_{weight}$) & 45\,s          \\
            No-face decay constant ($k_{decay}$) & 60\,s          \\
            Authentication threshold ($\tau$)    & 0.5            \\
            Initial trust ($X_0$)                & 1.0            \\
            Similarity percentile ($p$)          & 0.90           \\
            Enrollment frames per head direction & 5              \\
            \bottomrule
        \end{tabular}
    \end{table}

    \section{Evaluation}

    We conduct two studies to evaluate possible differences between platforms and usage conditions. The controlled study compares mobile and desktop performance under three scenarios (baseline, repeated head turns, and low light). The in-the-wild study evaluates naturalistic mobile usage, investigating how real-world behavior affects authentication performance.
    The studies involve \ParticipantCount{} volunteers aged \ParticipantAgeRange{} years, with a mean age of \ParticipantMeanAge{} years. The group includes \MaleParticipantCount{} male and \FemaleParticipantCount{} female participants recruited from personal and university networks. All participants completed enrollment and the controlled study with a fixed set of recordings. A subset subsequently contributed varying numbers of recordings to the in-the-wild study.

    \subsection{Enrollment}
    \label{sec:enrollment}

    Participants completed an enrollment session on both mobile and desktop devices. During enrollment, participants saw a blue dot moving around a circle on the device screen. They were instructed to follow the dot with their nose, causing their head to move through different orientations, including front, left, right, up, and down. Each participant completed the enrollment in both clockwise and counterclockwise directions to capture reference images from several head orientations.
    The approach then selects frames for each head direction. It samples 5 evenly spaced frames from a 2\,s window for each direction. Each enrollment results in 50 enrollment images per participant and device (5 directions $\times$ 2 enrollment rotation directions $\times$ 5 frames). Across all participants and devices, the study collects \EnrollmentRecordingCount{} enrollment recordings (\ParticipantCount{} participants $\times$ 2 devices $\times$ 2 enrollment rotation directions). Enrollment was performed under good lighting conditions and reused across all evaluation scenarios to reflect real-world situations where users typically enroll only once.

    \subsection{Controlled Study}
\begin{figure*}[tbh]
        \centering
        \includegraphics[width=0.9\textwidth]{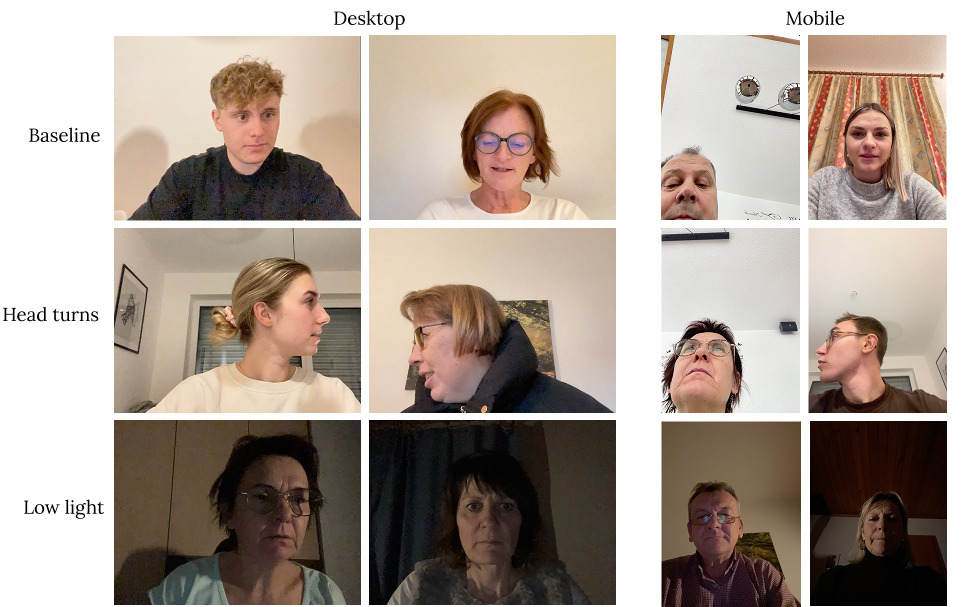}
        \caption{Controlled study examples on desktop and mobile. Rows show the baseline, repeated head turns, and low-light scenarios.}
        \label{fig:scenarios}
    \end{figure*}

    Participants were seated at a table and completed the recording tasks on an iPhone 12 Pro and a MacBook Pro 2021. An instructor explained each task, started the recordings, and kept the procedure consistent. Each recording lasted 3 minutes and captured the participant with the front-facing camera at 30\,Hz while they entered text on the device. Some recordings involved two participants in the same room. They completed the tasks independently and remained outside each other's camera view.
    The study included three scenarios that varied the recording conditions while retaining text entry as the common interaction task (Fig.~\ref{fig:scenarios}). Each participant completed all three scenarios on both mobile and desktop devices, which resulted in \ControlledStudyUsageRecordingCount{} controlled-study recordings (\ParticipantCount{} participants $\times$ 2 devices $\times$ 3 scenarios):

    Baseline scenario: Participants answered short reflective writing prompts under good indoor lighting. Their face was expected to remain mostly frontal towards the device, with head movements arising from reading, thinking, typing, or briefly looking away. This scenario served as the reference condition for comparison with the head-turn and low-light scenarios.

    Repeated head turns scenario: Three sheets with phrases were placed at different positions: one directly in front of the participant, one approximately 90 degrees to the right, and one approximately 90 degrees to the left. Participants were instructed to write down phrases from each sheet in sequence (front, left, right, front, \ldots), requiring them to rotate their heads. Lighting conditions remained the same as in the baseline scenario.

    Low light scenario: Participants answered questions similar to those in the baseline scenario, but under low light conditions. The room was darkened, and a single lamp was placed approximately 90 degrees to either the left or right side of the participant, creating strong shadows and overall poor facial illumination.
    
    \subsection{In-the-Wild Study}
\begin{figure*}[tbh]
        \centering
        \includegraphics[width=0.9\textwidth]{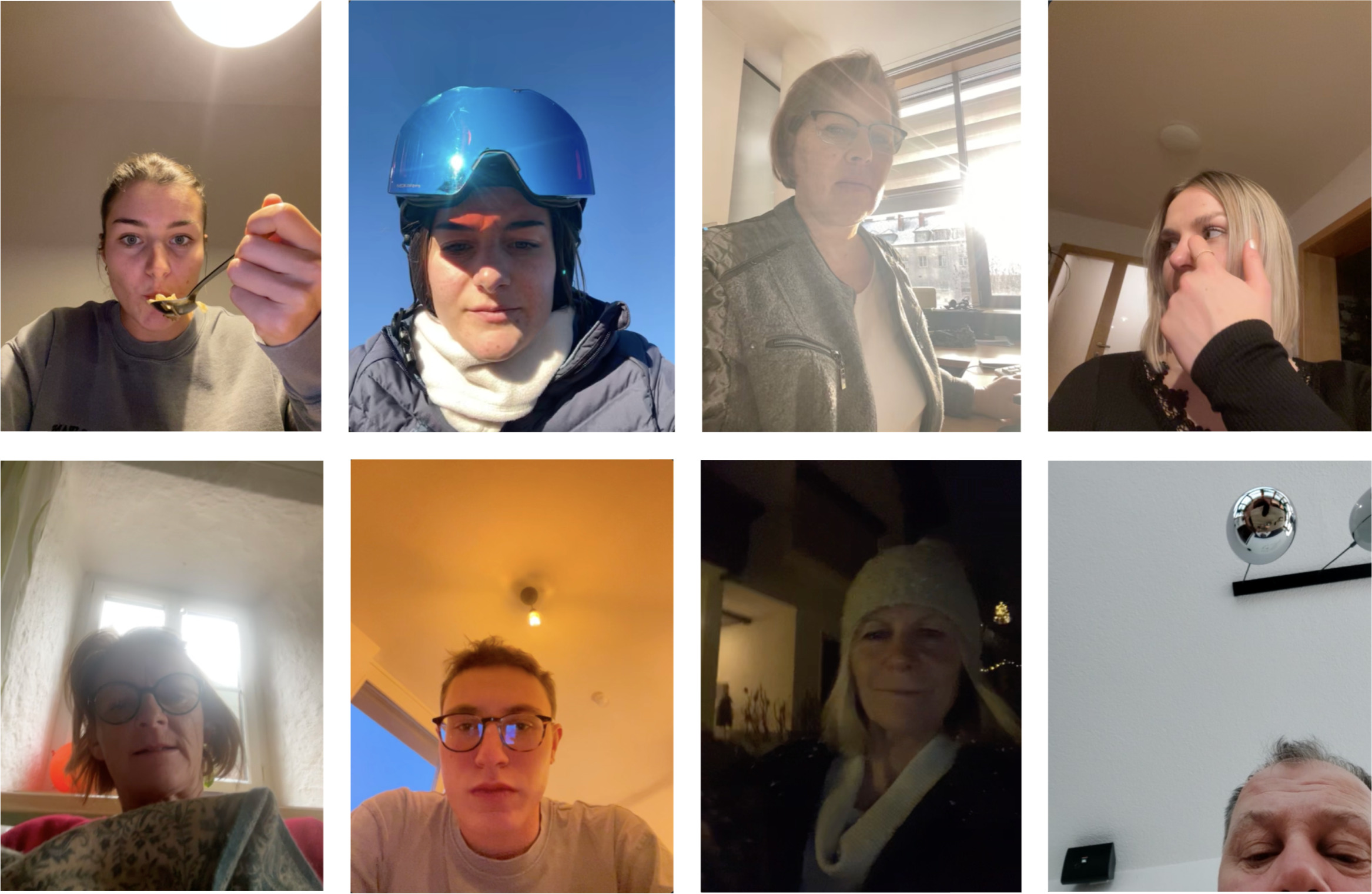}
        \caption{In-the-wild study examples with varying lighting, head positions, and occlusions.}
        \label{fig:in_the_wild}
    \end{figure*}

    After the controlled study, \InTheWildParticipantCount{} of the participants installed a custom application on their own smartphones and recorded in their everyday environments.
    \InTheWildAndroidParticipantCount{} used Android devices and \InTheWildIPhoneParticipantCount{} used iPhones.
    The instructor demonstrated the app and explained its features. Participants were asked to complete at least four 5-minute recordings over several days. To keep participants in the app during recording, the app provided links to games, digital newspapers, and other websites that opened in an in-app browser. Participants freely selected among these activities while the front-facing camera recorded their face at 30\,Hz (examples in Fig.~\ref{fig:in_the_wild}).
    Participants completed \MinInTheWildRecordings{}--\MaxInTheWildRecordings{} (mean \MeanInTheWildRecordings{}) usage recordings over a period of 7 days, resulting in \InTheWildUsageRecordingCount{} in-the-wild recordings. The recordings contained variation in lighting, head position, occlusion, and movement.

    \subsection{Evaluation Methodology}

    The evaluation processes the recorded videos offline at 1\,Hz. Processing continues until the end of each video even if the trust score falls below the lock threshold. This preserves the complete trust-score trajectory and allows the first lockout time to be identified. In a deployed approach, the device would lock when this threshold is crossed. To simulate device takeovers, we construct evaluation videos from recordings of different participants. Each evaluation video contains three consecutive segments.

    \begin{enumerate}
        \item A genuine-user segment showing the enrolled participant using the device. It lasts 3 minutes for the controlled study and 5 minutes for the in-the-wild study.
        \item A 3\,s transition segment containing black frames between the two users. These frames simulate the front-facing camera being briefly covered while the unauthorized user takes the device.
        \item An unauthorized-user segment showing another study participant using the device without attempting to avoid detection. The approach continues to use the genuine user's enrollment data, and the segment has the same duration as the genuine-user segment.
    \end{enumerate}
    For the controlled study, each genuine-user recording is paired with recordings of all \ControlledPairingsPerGenuineVideo{} other participants captured with the same device type and under the same scenario. This produces \ControlledComposedEvaluationVideoCount{} composed evaluation videos. For the in-the-wild study, each genuine-user recording is paired with \InTheWildPairingsPerGenuineVideo{} randomly selected recordings from other participants. A fixed random seed makes these pairings reproducible. This produces \InTheWildComposedEvaluationVideoCount{} composed evaluation videos.

    \subsubsection{Genuine User Metrics}
    The genuine-user analysis measures whether the approach keeps the device unlocked while the genuine user is present. Although a genuine-user recording occurs in several composed evaluation videos, it is counted only once when calculating the genuine-user metrics.
    False Rejection Rate (FRR) is the percentage of genuine-user recordings in which the trust score falls below the lock threshold. Genuine User Kickout Time (GKT) is the time from the beginning of the recording until the first false lockout and is reported only for affected recordings. Genuine User Trust (GT) is the mean trust score across the genuine-user recordings.

    \subsubsection{Unauthorized User Metrics}
    For unauthorized users, the analysis measures whether and when the approach locks after another participant appears. False Acceptance Rate (FAR) is the percentage of eligible evaluation videos in which the unauthorized user is never locked out. Unauthorized User Lockout Time (ULT) is the time between the first frame of the unauthorized-user segment and the first subsequent lockout.
    Evaluation videos in which a false lockout occurs during the genuine-user segment are excluded from FAR and ULT because the device would already be locked before the unauthorized user appears.

    \section{Results}

    \subsection{Platform Comparison}

    In general, differences between desktop and mobile authentication results are small (Table~\ref{tab:controlled-results}). Desktop produces slightly higher GT and correspondingly longer ULT values than mobile. The overall difference in GT is statistically significant ($W = 87.0$, $p = .042$), whereas the difference in ULT does not reach statistical significance ($W = 79.5$, $p = .075$). This GT difference may partly stem from the mismatch between enrollment and usage posture on mobile devices. Desktop participants maintained a consistent frontal posture during both enrollment and usage. In contrast, mobile participants often held the phone directly in front of their face during enrollment but lowered it to a more comfortable typing position during actual use. This shift in camera angle may have slightly reduced mobile GT values. When testing each scenario separately with Bonferroni correction, none of the device comparisons remain significant for either GT (baseline p = .0173, head-turn p = .230, low-light p = .287) or ULT (baseline p = .0358, head-turn p = .474, low-light p = .0716). Since the overall GT effect is small and the separate scenario comparisons are not significant, we conclude that device type alone has little impact.

    All unauthorized users are locked out on both desktop and mobile devices. The only two false lockouts occur on desktop under low light. One possible explanation is the visibly noisier output of the MacBook webcam under these conditions, while the iPhone applies stronger image processing.

    \begin{table*}[tbh]
        \centering
        \caption{Controlled-study results by device and scenario. Parentheses report raw FRR counts and 90\% percentile (P90) values for GKT and ULT.}
        \label{tab:controlled-results}
        \begin{tabular*}{\textwidth}{@{\extracolsep{\fill}}llccccc}
            \toprule
            \textbf{Platform} & \textbf{Scenario} & \textbf{FRR} & \textbf{FAR} & \shortstack{{\scriptsize\textbf{mean}}\\\textbf{GT}} & \shortstack{{\scriptsize\textbf{mean}}\\\textbf{GKT}} & \shortstack{{\scriptsize\textbf{mean}}\\\textbf{ULT}} \\
            \midrule
            Desktop & Baseline & 0.0\% & 0.0\% & 0.86 & -- & 22\,s {{\scriptsize (28\,s)}} \\
            & Head turns & 0.0\% & 0.0\% & 0.80 & -- & 15\,s {{\scriptsize (23\,s)}} \\
            & Low light & 8.0\% {{\scriptsize (2/25)}} & 0.0\% & 0.72 & 128\,s {{\scriptsize (149\,s)}} & 10\,s {{\scriptsize (16\,s)}} \\
            \midrule
            Mobile & Baseline & 0.0\% & 0.0\% & 0.82 & -- & 19\,s {{\scriptsize (26\,s)}} \\
            & Head turns & 0.0\% & 0.0\% & 0.78 & -- & 14\,s {{\scriptsize (22\,s)}} \\
            & Low light & 0.0\% & 0.0\% & 0.70 & -- & 8\,s {{\scriptsize (13\,s)}} \\
            \bottomrule
        \end{tabular*}
    \end{table*}

    \subsubsection{Effect of Scenario Conditions}

    In contrast to the small device-related differences, the scenario conditions produce larger differences in both GT and ULT, as shown in Table~\ref{tab:controlled-results}. The baseline scenario produces the highest GT because its lighting and mostly frontal viewing position closely match the enrollment conditions. Repeated head turns reduce GT because non-frontal views temporarily produce lower similarity scores. However, participants repeatedly return to a frontal position, allowing the trust score to recover. Low light produces the lowest GT because shadows and poor facial illumination affect the observations more consistently throughout the recording. It is also the only controlled condition that produces false lockouts.

\begin{figure*}[tbh]
        \centering
        \includegraphics[width=\linewidth]{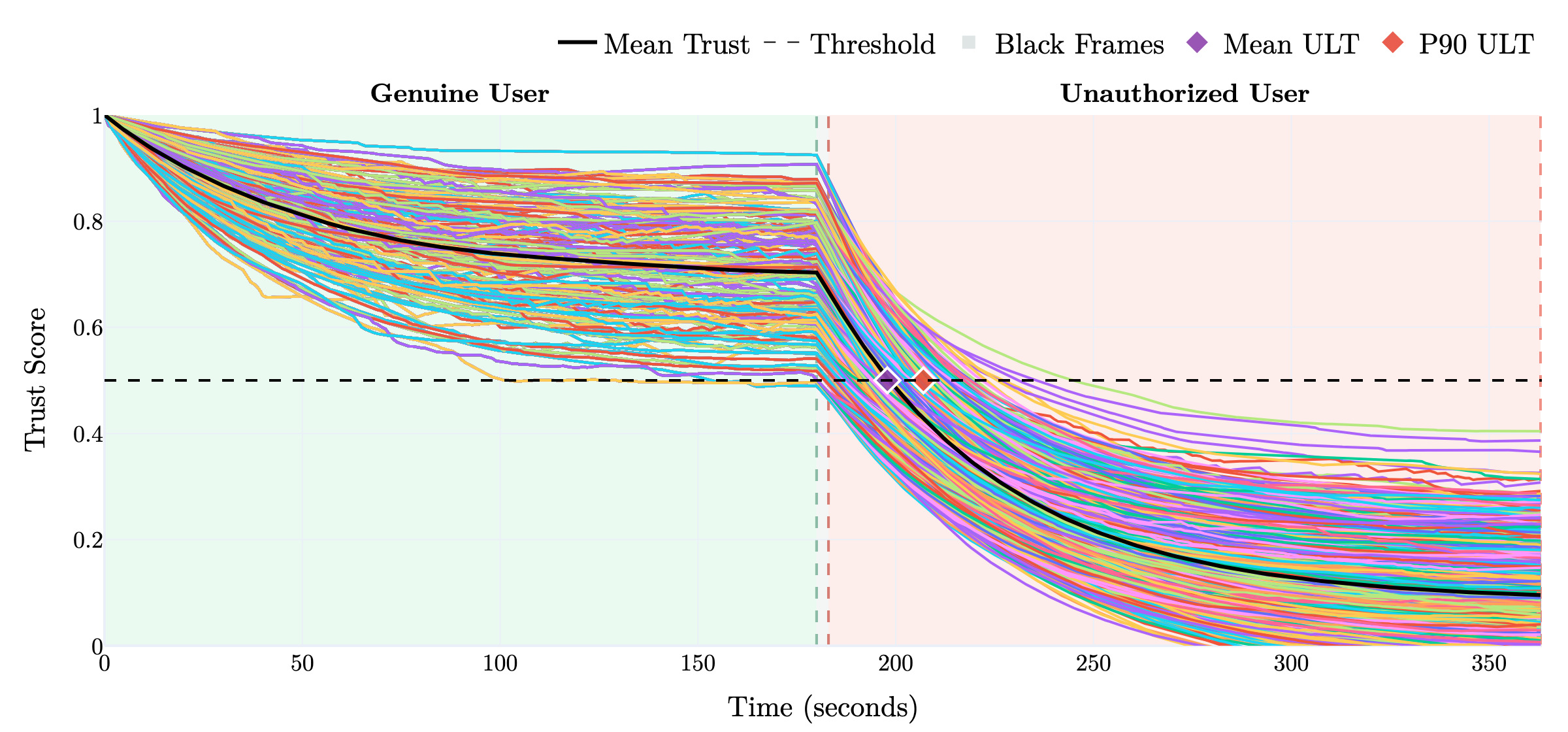}
        \caption{Controlled-study trust scores by scenario and device. Colored lines show composed videos, black lines show means, and the dashed line marks $\tau = 0.5$. Diamonds mark mean and P90 ULT.}
        \label{fig:trust-controlled}
    \end{figure*}

    Statistical tests confirm significant differences between the scenarios for both GT and ULT on both desktop and mobile devices (all $p < .001$). All scenario pairs also differ significantly after Bonferroni correction (all $p < .001$). Higher GT in the baseline scenario leaves more distance to the lock threshold and therefore results in longer ULT values after an unauthorized user appears. The lower trust scores under repeated head turns and low light begin closer to the threshold and result in progressively shorter ULT values. This inverse relationship between GT and ULT is visible throughout the evaluation results.
    Figure~\ref{fig:trust-controlled} illustrates the trust score traces
    across all composed videos, showing how the score evolves from the genuine user
    segment through the device handover into the unauthorized user segment.

    \subsection{In-the-Wild Study Results}

\begin{figure*}[tbh]
        \vspace{-8mm}
        \centering
        \includegraphics[width=\linewidth]{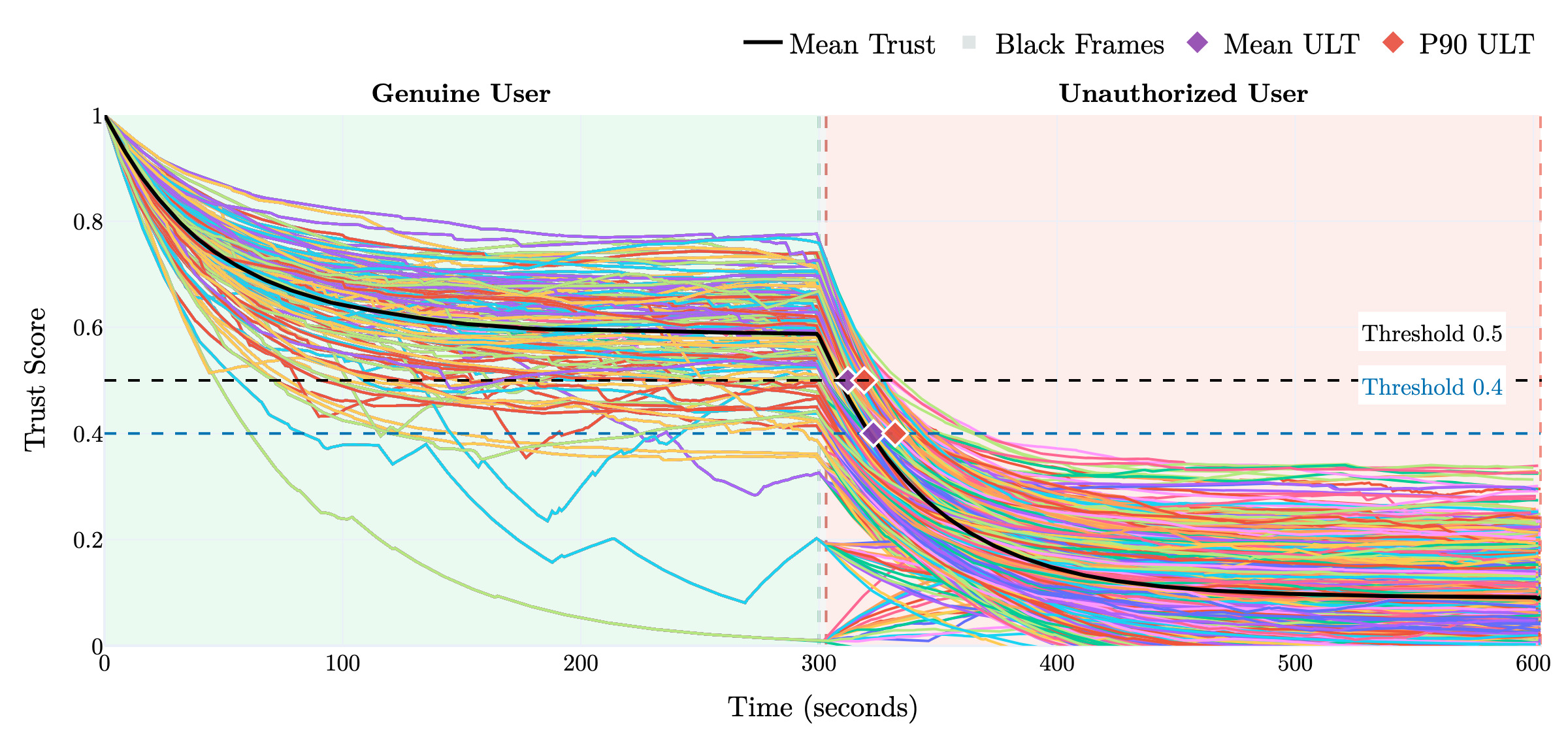}
        \caption{In-the-wild trust scores. Colored lines show composed videos, the black line shows the mean, and dashed lines mark $\tau = 0.5$ and $\tau = 0.4$. Diamonds mark mean and P90 ULT.}
        \label{fig:trust-wild}
    \end{figure*}

    The in-the-wild recordings produce lower and more variable trust scores than the controlled mobile recordings (Table~\ref{tab:wild-results}). At $\tau = 0.5$, mean GT decreases from 0.77 to 0.65, while FRR increases from 0.0\% to 22.3\%, corresponding to 27 of 121 genuine-user recordings. Figure~\ref{fig:trust-wild} shows that several genuine-user trajectories approach or cross the lock threshold during natural mobile use. FAR remains 0.0\%, and mean ULT decreases from 14\,s to 9\,s. This shorter ULT is consistent with the lower GT because the trust score starts closer to the threshold when the unauthorized user appears.
    To simulate users being able to configure the continuous authentication for either more security or more usability, we repeat the evaluation with a lower threshold of $\tau = 0.4$, which corresponds to more usability. FRR decreases to 8.3\%, corresponding to 10 of 121 recordings, while FAR remains 0.0\%. However, mean ULT increases from 9\,s to 20\,s. The appropriate
    threshold therefore depends on the desired balance between
    legitimate-user continuity and unauthorized-user lockout speed in a given deployment context.

    To complement these results, we conduct an additional descriptive analysis of the conditions present in the false-rejected recordings. Reduced facial visibility occurs in 21 of the 27 recordings with the $\tau = 0.5$ threshold. This includes the face being outside the camera view and partial occlusions caused by hands, eating or drinking, headwear, or the face being cut off by the frame. The analysis does not isolate causal effects, but shows that reduced visibility frequently accompanies false lockouts during natural mobile use, hence is a key contributor to such false lockouts.

    \begin{table*}[tbh]
        \centering
        \caption{In-the-wild results at two thresholds. Parentheses report raw FRR and FAR counts and P90 values for GKT and ULT.}
        \label{tab:wild-results}
        \begin{tabular*}{\textwidth}{@{\extracolsep{\fill}}lccccc}
            \toprule
            \textbf{Study} & \textbf{FRR} & \textbf{FAR} & \shortstack{{\scriptsize\textbf{mean}}\\\textbf{GT}} & \shortstack{{\scriptsize\textbf{mean}}\\\textbf{GKT}} & \shortstack{{\scriptsize\textbf{mean}}\\\textbf{ULT}} \\
            \midrule
            Controlled mobile $\tau = 0.5$ & 0.0\% & 0.0\% & 0.77 & -- & 14\,s {{\scriptsize (23\,s)}} \\
            In-the-wild $\tau = 0.5$ & 22.3\% {{\scriptsize (27/121)}} & 0.0\% & 0.65 & 133\,s {{\scriptsize (240\,s)}} & 9\,s {{\scriptsize (16\,s)}} \\
            In-the-wild $\tau = 0.4$ & 8.3\% {{\scriptsize (10/121)}} & 0.0\% & 0.65 & 149\,s {{\scriptsize (232\,s)}} & 20\,s {{\scriptsize (29\,s)}} \\
            \bottomrule
        \end{tabular*}
    \end{table*}

    \FloatBarrier

    \section{Conclusion}
    \label{sec:conclusion}

    This paper examined how continuous face authentication is affected by desktop and mobile platform choice and by different usage conditions.
    The controlled study shows only small differences between mobile and desktop devices. Overall, mean GT and ULT values differ only slightly between both platforms. The overall GT difference is statistically significant, but the ULT difference is not, and the separate scenario comparisons do not remain significant after correction. The small difference in GT may be explained by a stronger mismatch between enrollment posture and typing posture on mobile, where participants often lowered the phone during usage. All unauthorized users are locked out on both platforms. Both false lockouts occur on desktop in the low light scenario. The results therefore indicate that the observed authentication behavior of the approach is similar across mobile and desktop devices under the evaluated conditions.
    The controlled study also shows that the evaluated usage conditions have a stronger effect than device type. The baseline scenario produces the highest GT and the longest ULT values. The repeated head turns scenario reduces both metrics, but does not cause a false lockout. The low light scenario produces the lowest GT, the shortest ULT values, and both false lockouts. Scenario differences are statistically significant on both platforms. Low light therefore has the strongest measured effect among the controlled scenarios, while participants can recover from temporary head turns when they return to a frontal view.

    The in-the-wild study confirms that real-world behavior is more challenging for continuous face authentication. During natural device use, participants frequently looked away from the screen, held the device at low or changing angles, or used it in difficult lighting conditions. This reduced face visibility and caused gaps where the face was not observable, leading to more false lockouts for users.
    Using a lower authentication threshold reduces the number of false rejections, making the approach less interruptive during normal use~-- which comes at the cost of slower unauthorized-user lockout.
    In settings where rapid lockout is the main priority, faster rejection of unauthorized users may be preferred. In everyday personal device use, users might instead choose a lower threshold if it avoids unnecessary interruptions.
    Limitations of this work include limited study participant diversity, in-the-wild use only being covered on mobile devices, and exclusion of adversarial behaviors by unauthorized users. Future work includes addressing these limitations through larger in-the-wild studies across platforms, and investigating impact of adversarial behavior.
\begin{credits}
\subsubsection{\discintname}
%It is now necessary to declare any competing interests or to specifically
%state that the authors have no competing interests. Please place the
%statement with a bold run-in heading in small font size beneath the
%(optional) acknowledgments\footnote{If EquinOCS, our proceedings submission
%system, is used, then the disclaimer can be provided directly in the system.},
%for example: The authors have no competing interests to declare that are
%relevant to the content of this article. Or: Author A has received research
%grants from Company W. Author B has received a speaker honorarium from
%Company X and owns stock in Company Y. Author C is a member of committee Z.
The authors have no competing interests to declare that are relevant to the content of this paper.
Rainhard Dieter Findling is also employed at Google LLC.
\end{credits}
    \bibliographystyle{splncs04}
    \bibliography{references}

\end{document}